\documentclass[11pt]{article}
\usepackage{mymoriond,epsfig}
\usepackage{amsmath}
\usepackage{amssymb}

\def\Journal#1#2#3#4{{#1} {\bf #2}, #3 (#4)}

\def\NPB{{\em Nucl. Phys.} B}
\def\PLB{{\em Phys. Lett.}  B}
\def\PRL{\em Phys. Rev. Lett.}
\def\PRD{{\em Phys. Rev.} D}

\def\be{\begin{equation}}
\def\ee{\end{equation}}
\def\bea{\begin{eqnarray}}
\def\eea{\end{eqnarray}}

\begin{document}
\vspace*{4cm}

\title{PERTURBATIVE STABILITY OF THE QCD ANALYSIS OF DIS DATA}

\author{S.I.~ALEKHIN}

\address{Institute for High Energy Physics, 142281 Protvino, Russia}

\maketitle\abstracts{We perform pQCD analysis of  
the existing DIS data for charged leptons 
with account of corrections up to the NNLO.
The parton distributions, value of strong coupling constant,
and high-twist terms are extracted and their stability with respect to account 
of the NNLO corrections is analyzed. All the quantities are generally 
stable within their experimental errors.
Obtained value of the strong coupling constant is
$\alpha_s^{\rm NNLO}(M_{\rm Z})=0.1143\pm 0.0014 ({\rm exp})$ with a 
guess $\alpha_s^{\rm NNNLO}(M_{\rm Z})\sim 0.113$.}

Perturbative method is a powerful tool of the modern quantum field theory.
In particular, analysis of the deep-inelastic-scattering (DIS) 
data in terms of perturbative QCD (pQCD) 
allows for quantitative description of this process and extraction 
of parton distribution functions (PDFs) together with value of 
the strong coupling constant $\alpha_{\rm s}$, which can be used 
for calculation of the cross sections for
other processes with hadronic beams and targets
and check of self-consistency of Standard Model. 
The common practice for analysis of such kind
is to take into account only the $O(\alpha_{\rm s})$
(or next-to-leading-order (NLO)) corrections to the DIS cross sections
since the next-to-next-to-leading-order (NNLO) corrections has not been
completely calculated yet.
Meanwhile the value of $\alpha_{\rm s}$ is rather large at the values of 
transferred momentum $Q$ 
typical for existing DIS data and the higher-order (HO) 
correction may have impact on the value of extracted quantities.
Particular important is to know the HO PDFs that 
is motivated by the need to calculate
precise value of the Higgs boson production 
cross section since these calculations require account of the HO QCD 
corrections~\cite{Harlander:2002wh} and corresponding HO PDFs 
as consistent input.
Account of the HO corrections is also important for reducing the 
total uncertainty in the value of $\alpha_{\rm s}$ determined from 
the existing DIS data since the theoretical error due to HO 
corrections dominates the error in value of $\alpha_{\rm s}$ 
obtained in the NLO approximation~\cite{Alekhin:2000ch}.
Besides, account of HO corrections 
is necessary for clarification of nature of the $1/Q^2$
terms, which are apparent in the NLO analysis of DIS data. 
These terms correspond to the high-twist (HT) contributions, but 
also can be simulated by the HO contributions since they are 
proportional to factors $[\alpha_{\rm s}(Q)]^n$ and also fall with $Q$. 
If such simulation does take place,  
the HT terms observed in NLO should decrease with account 
of the HO corrections 
(see Ref.\cite{Kataev:1999bp} and references therein).

The NNLO corrections to the DIS coefficient functions 
have been calculated about decade ago, but 
substantial progress in calculation of the corrections
to the anomalous dimensions   
has been achieved only recently with estimation of their Mellin moments  
up to 14-th~\cite{Retey:2000nq}. This input
reduces uncertainties in the values of corresponding splitting functions 
up to O(\%) in the region of $x$
covered by existing DIS data  that would provide 
reasonable level of theoretical uncertainties due 
to incomplete knowledge of the NNLO evolution kernel in the analysis 
of these data~\cite{vanNeerven:2000wp}. 
We perform the QCD fit to the existing
data on DIS of charged leptons off proton/deuteron 
targets~\cite{data} with account of the available QCD
corrections to the coefficient and splitting functions up to the NNLO. 
Details of the analysis are published elsewhere~\cite{Alekhin:2001ih}.
We fitted PDFs, value of 
$\alpha_{\rm s}$ and the twist-4 contributions, which are included 
to the fitted structure functions $F_{\rm 2,L}$ in additive form: 
$F_{\rm 2,L}=F_{\rm 2,L}^{\rm LT,TMC}+{H_{\rm 2,L}(x)}/{Q^2},$
where $F_{\rm 2,L}^{\rm LT,TMC}$ are the leading-twist
terms with account of the target mass corrections
and $H(x)$ are parameterized in the piece-linear form 
with $x$-spacing equal to 0.1.
The principal feature of our analysis is quantitative estimation of the 
experimental errors in fitted parameters. These errors give 
natural scale 
for estimation of stability of the fit with respect to different
corrections:
Effect with the magnitude less than corresponding experimental error 
cannot be considered as significant since at this level
it can be simulated by fluctuation of the data.

\begin{figure}[h]
\centerline{\epsfig{file=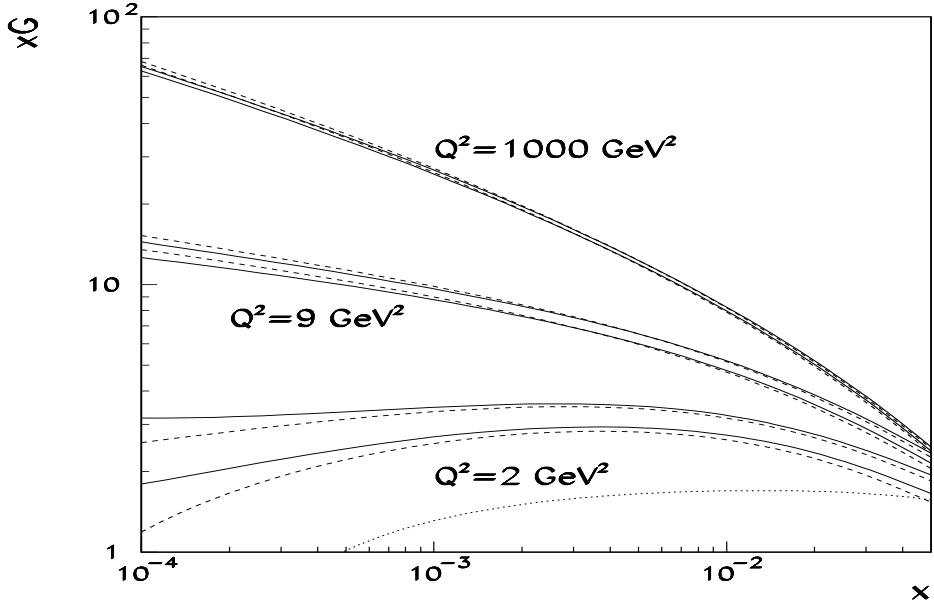,width=14cm,height=9.5cm}}
\caption{Experimental error bands 
for the gluon distributions obtained in our NNLO (full), 
NLO (dashes), and in the NNLO MRST analysis (dots).}
\label{fig:pdf}
\end{figure}

The NNLO PDFs obtained in our analysis
are generally comparable to the NLO ones within 
their experimental errors. Perturbative stability of the gluon 
distribution at small $x$ is demonstrated in Fig.\ref{fig:pdf}.  
For comparison we give in the same figure the gluon distribution 
obtained in the recent NNLO analysis of Ref.\cite{Martin:2002dr}.
The latter is smaller than ours and gets negative 
at $x\sim 10^{-4}$ at low $Q^2$, contrary to ours. 

\begin{table}[h]
\begin{center}
\caption{Values of $\alpha_{\rm s}(M_{\rm Z})$ obtained in different 
orders of pQCD.} 
\vspace{0.4cm}
\begin{tabular}{|c|c|}
\hline
LO&   $0.1301\pm0.0026 ({\rm exp})\pm0.0149 ({\rm RS})$ \\ \hline
NLO & $0.1171\pm0.0015 ({\rm exp})\pm0.0033 ({\rm RS})$ \\ \hline
NNLO & $0.1143\pm0.0014 ({\rm exp})\pm0.0009 ({\rm RS})$ \\ \hline
\end{tabular}
\end{center}
\label{tab:als}
\end{table}

\begin{figure}[h]
\centerline{\epsfig{file=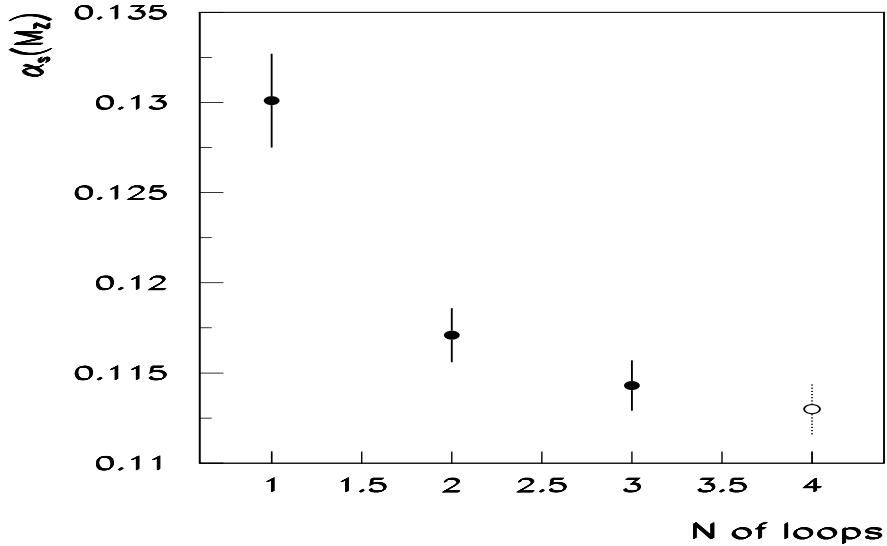,width=14cm,height=8cm}}
\caption{Values of $\alpha_{\rm s}(M_{\rm Z})$ and their experimental errors
obtained in different orders of pQCD (full symbols). Open symbol shows
extrapolation to the NNNLO.}
\label{fig:als}
\end{figure}

The values of $\alpha_{\rm s}$ obtained in different pQCD orders 
are given in Table~1. The associated renormalization scale 
(RS) error is calculated as the shift of $\alpha_{\rm s}$
under variation of the QCD RS 
from $Q^2$ to $4Q^2$. This shift is regularly considered as the 
uncertainty due to inaccount of the HO corrections. Such estimate  
is very crude since the the range of 
variation of the RS is conventional and no possible $x$-dependence
of the RS is taken into account. Nevertheless, comparing the RS  
errors to the corresponding change of $\alpha_{\rm s}(M_{\rm Z})$
with the orders of pQCD one can convince that they coincide with good accuracy.
Extrapolation of this regularity to the NNNLO 
illustrated in Fig.\ref{fig:als} leads to the estimate 
$\alpha^{\rm NNNLO}_{\rm s}(M_{\rm Z}) \sim 0.113$. 
The theoretical error in $\alpha_{\rm s}$ , which is dominated by 
the RS error is larger than the experimental error in the NLO. In the NNLO 
the RS error gets smaller than experimental one. Probably this 
is valid for the NNNLO too and thus both the central 
value and the error in $\alpha_{\rm s}$ extracted from the 
existing DIS data in NNNLO would be the about the same as in NNLO.

\begin{figure}[h]
\centerline{\epsfig{file=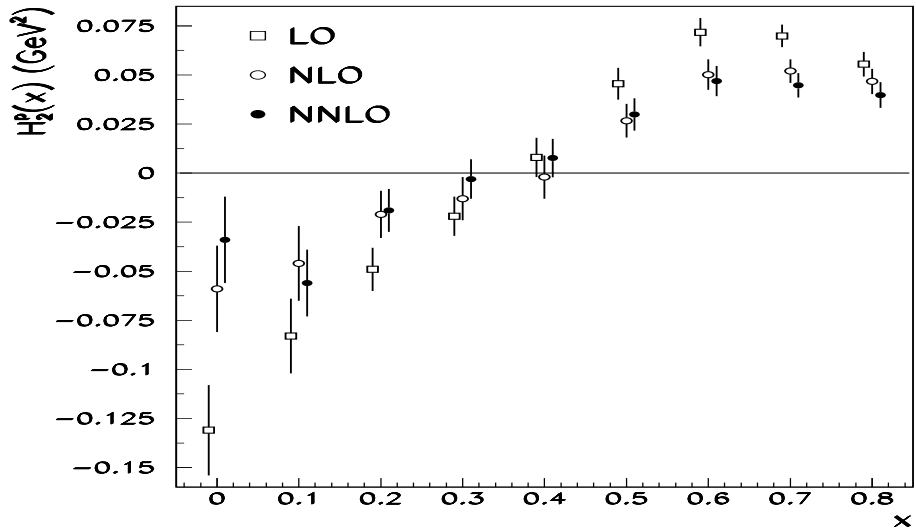,width=14cm,height=8cm}}
\caption{The twist-4 contribution to the 
proton structure function $F_2^p$ obtained in the different orders
of pQCD.}
\label{fig:ht}
\end{figure}

The HT terms are also stable with respect to the NNLO corrections 
(see Fig.\ref{fig:ht}). They do decrease from LO to NLO and from NLO to NNLO,  
but in the second case the shift is smaller and is comparable to  
the experimental error. The magnitude of the $H_2^p$
at $x=0.6$ and $Q^2=5~{\rm GeV}^2$ is $\sim 10\%$ of the LT term.
At lower $Q^2$ corresponding to the resonance region the 
relative contribution of the HT terms is even larger~\cite{Liuti:2001qk}. 
The value of $H_2^p$ is comparable to the experimental error in $F_2^p$ up to 
$Q^2\sim 20~{\rm GeV}^2$, i.e. an analysis 
of these data without account of the HT terms is biased
if one does not apply very stringent cut in $Q^2$. 
This result is in disagreement with the conclusion 
of Ref.\cite{Yang:1999xg} that the HT term in $F_2$ vanishes in the NNLO.
Meanwhile this disagreement may be insignificant since 
the analysis of Ref.\cite{Yang:1999xg} 
is based on the model-dependent determination of the HT terms 
and quality of the data description is poor
($\chi^2/{\rm NDP}$=1375/926 versus 2521/2274 in our fit).
Note also that the HT term in the nonsinglet part of $F_2^p$ 
obtained in the NNLO analysis of Ref.\cite{Schaefer:2001uh} does not vanish.
In the NNLO analysis of Ref.\cite{Kataev:1999bp} vanishing of 
the HT term in the neutrino-nucleon structure function $xF_3$
was observed, but in this case we see no disagreement with our results 
since error in $H_3$ obtained in Ref.\cite{Kataev:1999bp} 
is about order of magnitude larger than magnitude of $H_2$ 
obtained in our analysis. If the magnitudes of $H_2$ and $H_3$ are not very 
different, the conclusive study of the latter 
is possible only if it is determined with 
the precision $O(0.01)~{\rm GeV}^2$, which 
can be achieved in experiments with luminositities
typical for the proposed neutrino factories~\cite{Mangano:2001mj}.

In conclusion, we observe relative perturbative stability of the 
QCD analysis of existing data on DIS of charged leptons:
The change of PDFs, $\alpha_{\rm s}$, and HT terms
due to NNLO corrections is generally of the order of the 
experimental errors in these quantities.

\section*{Acknowledgments}
The work was supported by the RFBR grant 00-02-17432.

\end{document}